\documentclass[prl,twocolumn,showpacs]{revtex4}  
\usepackage{graphicx}  

\begin{document}


\title{The Resonance Scattering Phenomenon of Fast Negatively Charged Particles in a Single Crystal
\footnote{Nuclear Instruments and Methods, 1980, Vol. 170, No.1-3,  pp. 25-26} }

\author{G.\,V.\, Kovalev,  N.\,P.\, Kalashnikov \/\thanks}



\affiliation{Engineering-Physics Institute, Moscow }

\date{March, 1979}

\begin{abstract}
The energy spectrum of the extended attractive potential of a crystallographic row for negatively charged particles has quasi-bound  states. It follows that a negatively charged particle with small transversal momentum component ($p_{\bot} R <<1$) may undergo  resonance scattering. Thus the resonance scattering phenomenon can be observed in a single crystal, when fast electrons move with a small glancing angle ($\theta_0 << 1/pR$) to a crystallographic axis. The calculated results for the electrons and angular widths of
resonance peaks are consistent with experimental data.
\end{abstract}

\pacs{ 61.80.Fe}
\maketitle

Usually the resonance scattering phenomenon in 
elastic collisions is supposed to take place for slow 
particles, where $pR << 1$ ($p$ is the incident particle 
momentum and $R$ is the potential radius). In this 
work it is shown that fast ($pR >> 1$) negatively 
charged particles scattered by the extended attractive 
potential of a crystallographic row exhibit the resonance nature. This happens because the energy spectrum of the transversal potential of a crystallographic 
row has quasi-bound states. The importance of the 
 quasi-bound states was pointed out in the first article on 
the scattering of fast electrons in a single crystal [1]. 
The next works [2 - 4] contained efforts to develop this
idea.

However only in [5] it was observed that the 
scattering process has a resonance nature. The observation condition of resonance scattering in this 
case consists in the slow motion for the transverse component of incident particles ($p_{\bot} R <<1$, where $p_{\bot} = p \theta_0$, with $ \theta_0$ the glancing angle with respect to the 
cristallographic axis).

The wave function of a fast particle satisfies the Dirac equation

\begin{eqnarray}
(\Delta+E^2-M^2- 2EU)\Psi=0,
\label{r1}
\end{eqnarray}
where $M, E$ are the mass and energy of the particle, $U$ is  external potential ($E>> U$). For simplicity spin 
effects will be neglected. The exact wave function of particle in the continuum potential of a crystallographic row with longitudinal length $L$ and transverse radius $R$ ( $L > R $) is
\begin{eqnarray}
 \Psi= \left \{\begin{array}{ll} \Psi_1, &  \; -\infty < z \leq 0, \\
 \Psi_2, &  \; 0 \leq z \leq L, \\  
  \Psi_3, &  \; L \leq z < \infty.
\end{array} \right. 
\label{r2}
\end{eqnarray}
$\Psi_1, \Psi_3$ are the wave functions outside the interaction 
field which consist of a plane wave and a spherical 
divergent wave. $\Psi_2$ is the wave function inside the 
interaction field, which we decompose in the set of 
eigenfunctions in the potential $U$.   It has been 
assumed that the potential $U$ does not depend on the 
coordinate $z$ [ $U \equiv U(\vec{\rho})$] and the direction of the 
incident particle momentum is nearly parallel to the 
axis $z$.  Then  $\Psi_2$ can be written
\begin{eqnarray}
 \Psi_2 =\sum_{\alpha, m}Q_{\vec{p}_{\perp } }(\alpha, m) e^{i m \phi} Z_{\alpha, m}(\rho) e^{i z\sqrt{p^2-\alpha} }, 
\label{r3}
\end{eqnarray}
where $exp(im \phi) Z_{\alpha, m}(\rho)$ is the eigenfunction of the 
particle transverse motion, which satisfies the equation
\begin{eqnarray}
\left[\frac{1}{\rho} \frac{\partial }{\partial \rho} \rho \frac{\partial }{\partial \rho} +\frac{1}{\rho^2}\frac{\partial^2}{\partial \phi^2}- 2EU(\rho)\right]  e^{i m \phi} Z_{\alpha, m}(\rho) =\nonumber \\
-\alpha  e^{i m \phi} Z_{\alpha, m}(\rho) .
\label{r4}
\end{eqnarray}
Using the asym ptotic behaviour for the wave func­
tion (2) and the continuity conditions on the 
boundaries $Z =0$ and $Z=L$, we obtain the scattering 
amplitude [6]
\begin{eqnarray}
f(\theta, \phi)=
\frac{p}{2 \pi i} \sum Q_{\vec{p}_{\perp i} }(\alpha, m) Q^{*}_{\vec{p}_{\perp f} }(\alpha, m) \\ \nonumber 
 [ \exp(i \frac{\alpha-p_{\perp }^{2}}{2 p}L) -1],
\label{r5a}
\end{eqnarray}
where $\vec{p}_{\perp i}$ and $\vec{p}_{\perp f}$ are the transverse components of 
the initial and final momenta,
\begin{eqnarray}
Q_{p_{\perp } }(\alpha, m) =\int d^2 \, \vec{\rho}\, Z_{\alpha, m} ( \vec{\rho} ) \exp(i \vec{p}_{\perp } \vec{\rho} ).
\label{r5b}
\end{eqnarray}
One can show that the coefficient $Q_{p_{\perp } }(\alpha, m) $ in an 
arbitrary potential is
\begin{eqnarray}
Q_{p_{\perp } }(\alpha, m) =\delta( \vec{p}_{\perp } - \vec{p}_{\alpha} ) - \frac{ f(\vec{p}_{\perp} ,\alpha,m) }{p^2_{\perp} -p^2_{\perp} -i \delta},\nonumber \\
p^2_{\alpha} = \alpha   \; \;\; \text{for } \; \;\;   \alpha >0,
\label{r5c}
\end{eqnarray}
and
\begin{eqnarray}
Q_{p_{\perp } }(\alpha, m) =- \vec{p}_{\alpha} ) - \frac{ f(\vec{p}_{\perp} ,\alpha,m) }{p^2_{\perp} + p^2_{\perp} },\nonumber \\
p^2_{\alpha} = - \alpha  \; \;\; \text{for } \; \;\;     \alpha < 0.
\label{r5d}
\end{eqnarray}
Here the function $ f(\vec{p}_{\perp} ,\alpha,m) $  is given by
\begin{eqnarray}
f(\vec{p}_{\perp} ,\alpha,m)= 2E
\int d^2 \, \vec{\rho}\,  e^{i \vec{p}_{\perp } \vec{\rho} } \, U(\vec{\rho})  e^{-i m \phi }  Z_{\alpha, m} ( \vec{\rho} ) .
\label{r5f}
\end{eqnarray}

If $\alpha = p^2_{\perp} $, the function $f(\vec{p}_{\perp} ,\alpha,m)$ is the two-dimensional scattering amplitude by the potential $U(\rho)$.   
Using the optical theorem   $\sigma_{tot}  = (4 \pi /p) Im[ f(\theta_0, \phi_0)]$, 
we have
\begin{eqnarray}
\sigma_{tot}=4 \sum_{\alpha,m} \left| \frac{f(\vec{p}_{\perp i} ,\alpha,m)}{p^2_{\perp i}-\alpha}\right|^2 \sin^2 (\frac{p_{\perp i}^{2}-\alpha}{4 p}L) .
\label{r5h}
\end{eqnarray}

As  $f(\vec{p}_{\perp i} ,\alpha,m)$ has at resonance nature in an attractive 
potential for slow transverse motion ($p_{\perp i} <<1 $), the 
total cross section of the fast particles is represented 
by the Breit —Wigner formula [5]
\begin{eqnarray}
\sigma \simeq \pi  \frac{L}{p} \frac{\frac{1}{4}  \Gamma^2 }{(\theta_0-\theta_{0 res})^2+\frac{1}{4} \Gamma^2},
\label{r6a}
\end{eqnarray}
where
\begin{eqnarray}
\theta_{0 \, res} \simeq \left(\frac{\pi^2 n^2+m^2}{ E^2 R^2} -2 \frac{ V_0}{E}\right)^{1/2} \label{r6b1} \\
\Gamma  \simeq \theta_{0 \, res} \cdot  \exp- |2E V_0 R^2 -\pi^2 n^2|^{1/2}.
\label{r6b2}
\end{eqnarray}

\begin{figure}
\centering
		\includegraphics[width=0.45\textwidth]{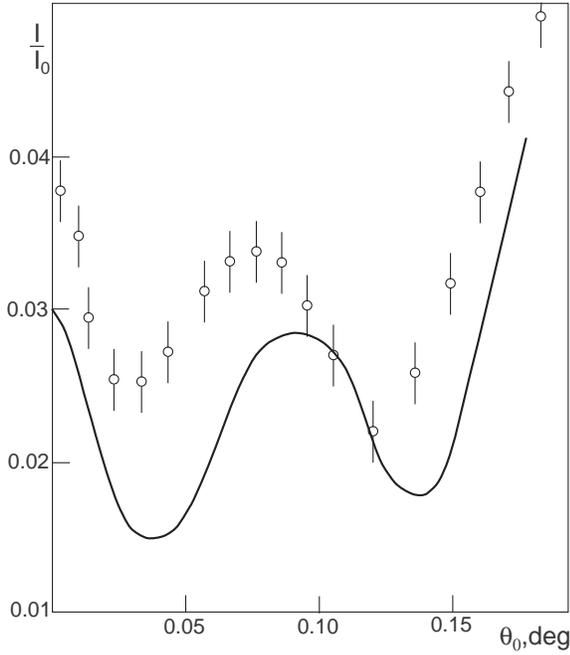}
\caption{The solid curve is the relative intensity of the 
scattered electrons ($E = 15 MeV$) by a single crystal $Si $ ($L= 
1.4\mu$ ) depending on the glancing angle $theta_0$ with respect 
the crystallographic axis (111). The experimental data a 
taken from [7].}
	\label{fig:fig_2}
\end{figure}

In fig. 1 the dependence of the scattering of $15$MeV 
electrons in a silicon crystal with thickness $1.4\mu$ upon the tilt angle is shown. For silicon the potential  
parameters are $U_0 = 23 eV$,  $R^{-1} = m e^2 Z^{1/3} = 9.0 * 10^{3}$eV and we have the narrow central peak $\theta_{0} \simeq 0.02^{o}$ [5].   For the lateral peak we get $\theta_{0 res} \simeq 0.1^{o}$  
and $\Gamma \simeq 0.5*\theta_{0 res} $. Thus the solid curve gives satisfactory agreement with the experimental data (see Fig. 1).


\begin{thebibliography}{10}

\bibitem{KSHH}
H.~J. Kreiner, F. Bell, R. Sizmann, D. Harder and  W. Huttl. -
{\em Phys. Lett.}, {\bf 33A}, 135 (1970).

\bibitem{Kumm_1972-1}
H. Kumm et al.,{\em Rad. Effects }, {\bf 12}, 53 (1972).

\bibitem{BKV}
A.~Ya. Bobudaev, V.V. Kaplin and S.A. Vorobev,  -
{\em Phys. Lett.}, {\bf 45A}, 71 (1973).

\bibitem{KF}
K. Komaki, F. Fujimoto ,  -
{\em Phys. Lett.}, {\bf 49A}, 445 (1974).


\bibitem{Kov-Kal_1979-1}
N.~P. Kalashnikov and G.~V. Kovalev,{\em JETP Lett.}, {\bf 29}, 302 (1979).


\bibitem{Kalashnikov-Strikhanov_1975-1}
N.~P. Kalashnikov and M.~N. Strikhanov,{\em Nuovo Cim.}, {\bf 29}, 9 (1975).


\bibitem{Schiebel-Worm_1976}
U.~Schiebel, E.Worm. {\em {Phys. Lett., }}, {\bf 58A}, 252 (1976).



\end{thebibliography}

\end{document}